\input{aipcheck}

\documentclass[
    ,final            
  ]
  {aipproc}

\layoutstyle{6x9}

\begin{document}

\title{New measurement of charge asymmetry $x{F}_3$ from HERA}


\author{L. Schoeffel}{
  address={CEA Saclay, Irfu/SPP, 91191 Gif-sur-Yvette Cedex, FRANCE}
}

\begin{abstract}
After presenting the recent measurements of neutral current cross section
in DIS at HERA, we explain  the effect of the
$\gamma-Z_0$ interference at the electro-weak scale, visible on these data.
Then, the beam charge difference $x{F}_3$ is measured
and the interference itself is extracted.
Results are discussed in the context of perturbative QCD.
\end{abstract}

\maketitle


\section{Introduction}

\begin{figure}[ht]
  \includegraphics[height=0.5\textheight]{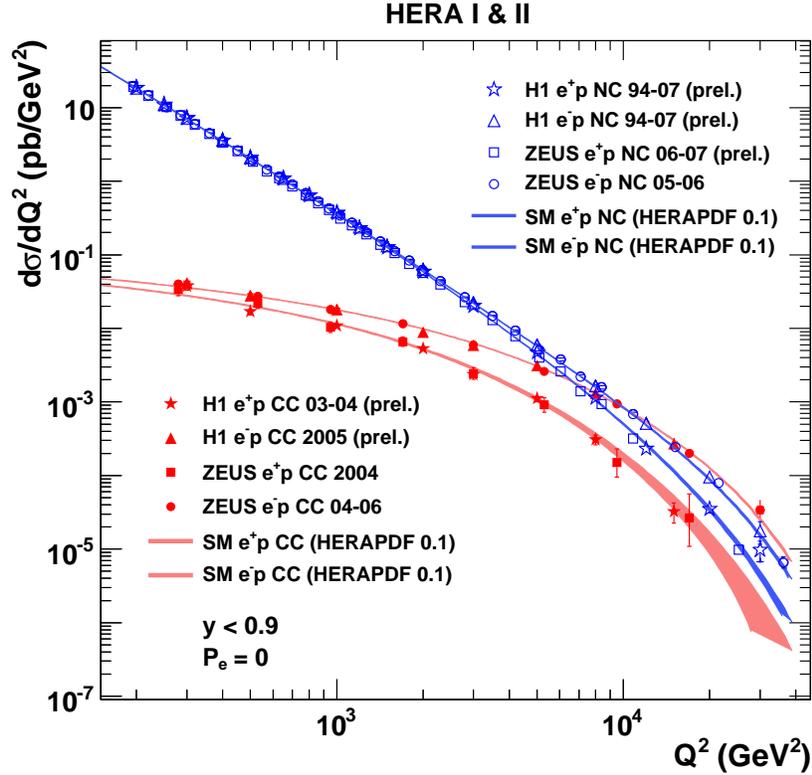}
  \caption{Neutral Current (NC) and Charge Current (CC) cross sections in DIS
  ($Q^2$ dependence) 
  as measured by
  H1 and ZEUS experiments at HERA. 
  Cross sections obtained from electron-proton and positron-proton collisions are 
  shown (see Ref. \cite{klein} for a complete presentation of these results).}
  \label{nccc}
\end{figure}

HERA was a collider, located at the DESY laboratory in Hamburg,
with the capablility to scatter (polarised) electrons and positrons
off protons, at a center of mass energy, $\sqrt{s}$, 
of about 320\,GeV. The first HERA data were taken in summer 1992.
It ceased operations in June 2007. 
The total luminosity collected by both experiments
H1 and ZEUS is  $\sim 0.9$ fb$^{-1}$.
The core results of HERA 
(with H1 and ZEUS experiments)
are related to deep inelastic lepton-proton scattering (DIS),
where
the proton structure is probed by a virtual photon ($\gamma^*$), or weak
bosons ($Z_0$ or $W^{\pm}$).   
The typical resolution scale is then inversely proportional to the
 momentum of the exchanged boson $\sim 1/\sqrt{Q^2+M_{Z,W}^2}$.  
Main results concerning measurements of DIS cross sections  
are summarised in Fig. \ref{nccc} \cite{klein,hz}.
The Neutral Current (NC) and Charge Current (CC) cross sections in DIS
are displayed. A NC process corresponds to the reaction $ep \rightarrow eX$,
where the incoming lepton is scattered with a change in momentum. Then,
 the exchanged boson is a (neutral) virtual photon at low $Q^2$
 and a virtual photon or a $Z_0$ boson at large $Q^2$ ($Q^2>M_Z^2$).
In a CC process, the lepton is scattered as a neutrino, $ep \rightarrow \nu X$,
which means that the exchanged boson is charged, namely $W^{\pm}$.
That's why values of CC cross sections are a few orders of magnitude below the
NC ones at low $Q^2$ and become of comparable size at large $Q^2$,
when $Q^2$ is above $M_Z^2$ or $M_W^2$. This is a fundamental observation,
visible in Fig. \ref{nccc}. This corresponds exactly to the Electro-Weak (EW)
unification in the Standard Model.

\section{Experimental results}

In Fig. \ref{nccc}, we also present cross section measurements for
$e^+p$ and $e^-p$ collisions separately. For CC processes, the difference between
 both lepton beam charges is obvious. In $e^+p$ and $e^-p$ collisions, $W^+$ 
 or $W^-$ bosons
 are exchanged (respectively).
 Then, the reactions at the proton vertex do not probe the same
 parton densities.
 Quarks of types $u$ and $d$ are probed with  $W^-$ boson
 and quarks of types ${\bar u}$ and
${\bar d}$ in the case of  $W^+$ boson exchange.

For NC cross sections, we also observe a difference with the
lepton beam charges. 
This last observation is less trivial to explain.
At large $Q^2$, for $Q^2 > M_Z^2$,
measured values of 
$e^-p \rightarrow e^-X$ (NC cross section) are greater 
than values for $e^+p \rightarrow e^+X$. 
And the
effect is more and more sizable when increasing $Q^2$.
This effect of beam charge asymmetry for measurements of NC cross section
at large $Q^2$ is resulting from the interference between 
interactions mediated by 
virtual photon and  $Z_0$ boson. Indeed, at large
$Q^2$, both are contributing to the production of NC events with the same
final state. Therefore they can interfere and the effect of this
interference is producing the different values
observed for $e^+p$ and $e^-p$ values of NC cross section. We define 
the function $x\tilde{F}_3$
as follows

\begin{equation}
\frac{d^2\sigma^-_{NC}}{dx dQ^2} 
-\frac{d^2\sigma^+_{NC}}{dx dQ^2} 
= \frac{2 \pi \alpha^2}{x Q^4}
[2 x\tilde{F}_3],
\label{main}
\end{equation}
with
\begin{equation} 
x\tilde{F}_3 \simeq - a_e \frac{\kappa_w Q^2}{Q^2+M_Z^2} xF_3^{\gamma Z}.
\label{eq2}
\end{equation}  
In this expression, $\kappa_w^{-1} = 4 M_W^2/M_Z^2~(1 - M_W^2/M_Z^2)$. Parameters
 $v_e$ and $a_e$ are the vector and 
axial vector couplings of the lepton to the $Z^0$ boson. They are related to 
the weak isospin of the lepton. Namely, $v_e=-1/2+2 \sin^2\theta_w$ and 
$a_e=-1/2$ where $\theta_w$ is the electroweak mixing angle. 

\begin{figure}[htbp]
\centering
  \includegraphics[height=0.4\textheight]{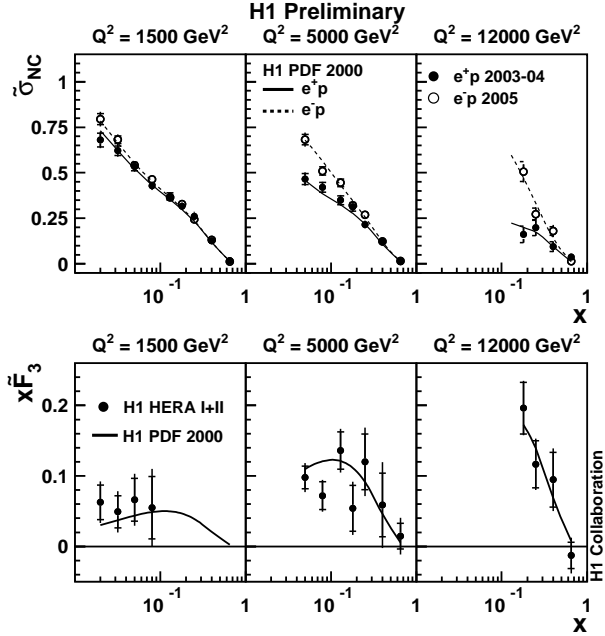}
  \caption{$x\tilde{F}_3$ extracted from  
  measurements of NC DIS cross section
  at large $Q^2$ (see text).}\label{xf3h1}
\end{figure}

\begin{figure}[htbp]
\centering
  \includegraphics[height=0.5\textheight]{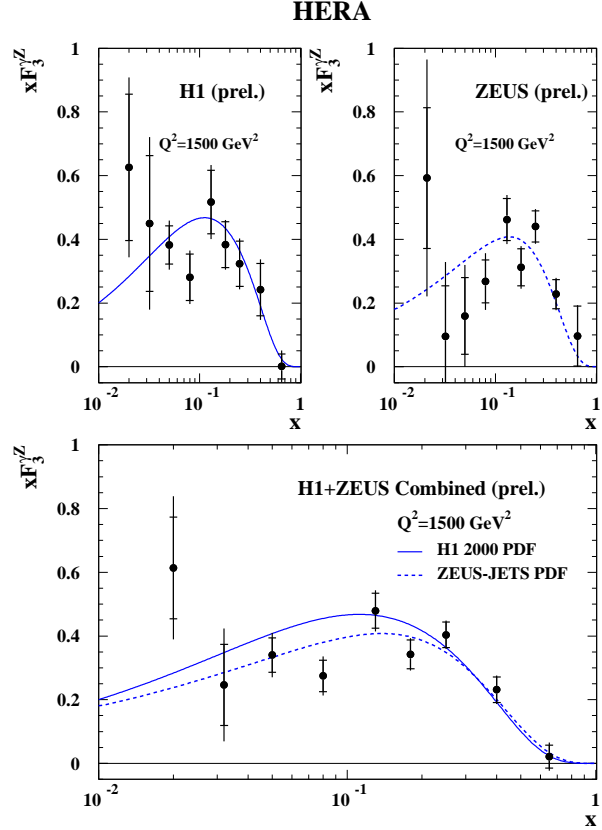}
  \caption{The interference term $xF_3^{\gamma Z}$ (see Eq. \ref{eq2})
  extracted by H1 and ZEUS experiments, and the combined
  values. 
}\label{xf3comb}
\end{figure}

Of course, measurements of NC cross section, illustrated in
Fig. \ref{nccc} in bins of $Q^2$, can be done more
differentially in Bjorken $x$ and $Q^2$. This is shown in
 Fig. \ref{xf3h1} for large $Q^2$, when the effect of
 $x\tilde{F}_3$ is sizeable \cite{klein, hz}. From the difference between
 $e^-p$ and $e^+p$ measurements, we can
 extract directly the function $x\tilde{F}_3$, using 
 its definition in Eq. (\ref{main}) (see Fig. \ref{xf3h1}).
 As discussed above, 
this is a direct measurement of the $\gamma-Z_0$ interference and its
impact on NC cross section at the EW scale is observed.
In this case, HERA is used as an interferometer at the EW scale.
This is obviously a highly non trivial experimental issue.
From Eq.
(\ref{eq2}), we can then compute directly the interference term,
apart from EW propagators, $xF_3^{\gamma Z}$,
which gives the magnitude of this interference.
Results are presented in Fig. \ref{xf3comb}.

\section{Discussion and outlook}

Finally, we can discuss the interest of this beam charge
asymmetry at the EW scale in the context of parton densities.
The main interest of  $xF_3^{\gamma Z}$  comes from its
expression in terms of quarks densities
\begin{equation}
xF_3^{\gamma Z}\simeq 2u_v + d_v
\label{epxf3}
\end{equation}
where $u_v$ and $d_v$ are the valence distributionsfor
$u$ and $d$ quarks.
In global fits of parton densities, valence type disctributions
are essentially constrained by  fixed target experiments 
(BCDMS and NMC) in the kinematic domain 
$x<0.2$ and $Q^2<150$ GeV$^2$.
With $x\tilde{F}_3$ measurements, we get the valence distribution
$2u_v + d_v$ (and its dependence in $x$) at very large $Q^2$,
above the EW scale. This provides reference data points for 
valence distributions that complements their determination
from fixed target experiments.
The statistical uncertainty on $x\tilde{F}_3$ is still large
(see Fig. \ref{xf3h1}).
However, as displayed in Fig. \ref{xf3comb}, a combination of 
H1 and ZEUS results is of great interest in order to reduce errors.
The present result is still premilinary, with only a part
of the statistics on tape analysed \cite{hz}. Therefore, in a near future, we expect
more accurate measurements of $x\tilde{F}_3$ and a greater
impact on PDFs determination.


\end{document}